\documentclass[journal]{IEEEtran}
\usepackage{graphicx}
\usepackage{subfigure}
\usepackage{algorithm}
\usepackage{algorithmic}
\usepackage{cite}
\usepackage{amsfonts}
\usepackage{psfrag}
\usepackage{footnote}
\usepackage{amsmath}
\usepackage[usenames]{color}
\usepackage{colortbl}
\usepackage[switch,pagewise]{lineno}

\newtheorem{lemma}{\textbf{Lemma}}
\newtheorem{theorem}{\textbf{Theorem}}

\newcommand{\rb}[1]{\raisebox{1.2ex}[0pt]{#1}}

\begin{document}
%
%
\title{AWG-based Non-blocking Clos Networks}

\author{Tong Ye,~\IEEEmembership{Member,~IEEE,}
        Tony T. Lee,~\IEEEmembership{Fellow,~IEEE,}
        and Weisheng Hu,~\IEEEmembership{Member,~IEEE}
\thanks{This work was supported by the National Science Foundation of China (61271215, 61172065, and 60825103),
973 program (2010CB328205, 2010CB328204), and Shanghai 09XD1402200. }
\thanks{The authors are with the State Key Laboratory of Advanced Optical Communication
Systems and Networks, Shanghai Jiao Tong University, Shanghai
200030, China. (e-mail: \{yetong, ttlee, wshu\}@sjtu.edu.cn}
}


\maketitle

\begin{abstract}
The three-stage Clos networks remain the most popular solution to
many practical switching systems to date. The aim of this paper is
to show that the modular structure of Clos networks is invariant
with respect to the technological changes. Due to the wavelength
routing property of arrayed-waveguide gratings (AWGs), non-blocking
and contention-free wavelength-division-multiplexing (WDM) switches
require that two calls carried by the same wavelength must be
connected by separated links; otherwise, they must be carried by
different wavelengths. Thus, in addition to the non-blocking
condition, the challenge of the design of AWG-based multistage
switching networks is to scale down the wavelength granularity and
to reduce the conversion range of tunable wavelength converters
(TWCs). We devise a logic scheme to partition the WDM switch network
into wavelength autonomous cells, and show that the wavelength
scalability problem can be solved by recursively reusing similar,
but smaller, set of wavelengths in different cells. Furthermore, we
prove that the rearrangeably non-blocking (RNB) condition and route
assignments in these AWG-based three-stage networks are consistent
with that of classical Clos networks. Thus, the optimal AWG-based
non-blocking Clos networks also can achieve 100\% utilization when
all input and output wavelength channels are busy.
\end{abstract}


\IEEEpeerreviewmaketitle

\section{Introduction}\label{intro}
\IEEEPARstart{A}{long} with rapid developments in optoelectronic
technologies in recent years, wavelength-division-multiplexing (WDM)
optical networks have emerged in response to the exponentially
growing demand for network capacity. Research in high-speed WDM
switch technology is inspired by the immense transmission bandwidth
offered by the WDM networks: experiments have demonstrated that an
individual optical fiber can carry more than 96 wavelength channels,
and each channel can support a data transmission rate up to 100
Gbit/s \cite{Salsi:OFC2011}. Furthermore, considerable progress has
been made in photonic devices, among which the combination of the
arrayed-waveguide grating (AWG) and the tunable wavelength
converters (TWCs) is considered a promising candidate to synthesize
WDM switches.

The AWG is a passive optical component with a low insertion loss.
Using wavelength routing, the AWG can forward the signal from an
input to an output without any contention \cite{Ye:JLT2012}. The
TWC, on the other hand, can convert an input wavelength to any of
the output wavelengths in the conversion range. The results reported
in \cite{Blumenthal:JSTQE,Liu:JLT2006,Liu:JLT2007} demonstrate that
the TWC can perform wavelength conversion at speeds of 40 Gb/s, 160
Gb/s and 320 Gb/s. In conjunction with the TWCs, the AWG can perform
high-speed packet switching, as demonstrated by the integrated
$8\times8$ AWG-based switching system described in
\cite{Blumenthal:JSTQE}.

Scalability is the key issue that challenges the design of
large-scale AWG-based WDM switches. First, the AWG with a large port
count suffers from serious coherent crosstalk if the same wavelength
is simultaneously fed into more than 15 inputs
\cite{Gaudino:ICC2008,Bianco:TON}. This limitation is difficult to
resolve through physical design \cite{Mikhailov:OFC2003}. Second,
the cost sharply increases with the conversion range of TWCs in
practice \cite{Pattavina:INFOCOM2006}. Third, the wavelengths should
be fully utilized because wavelengths are the precious resource in
the optical communication window \cite{Weichenberg:JOCN2009}. In
this paper, the size of the wavelength set associated with the
switch is referred to as \emph{wavelength granularity}.

Scalable large switching networks are usually made by combining
multiple stages of small-sized modular switches. The class of
non-blocking multistage switching networks was first proposed in
1953 by Charles Clos in his seminal paper \cite{Clos:BSTJ1953}. The
original three-stage Clos networks remain the most popular solution
to many practical switching systems to date. In spite of the rapid
developments in integrated circuit and photonic technologies during
the last several decades, the theory of Clos networks retains its
relevance. The most important characteristic of a Clos network is
its modular structure, which is invariant with respect to the
technological changes.

In an optimally designed rearrangeably non-blocking (RNB) Clos
network, all modular switches and interconnection links are fully
utilized when all inputs and outputs are loaded. As for AWG-based
Clos networks, the non-blocking condition requires that any two
connections from inputs to outputs be either space-interleaved or
wavelength-interleaved, i.e., two connections carried by the same
wavelength must be connected by separated links, otherwise they must
be carried by different wavelengths. To be consistent with the
original non-blocking theory of Clos networks, another key issue for
constructing AWG-based non-blocking Clos networks is the optimality
criterion of network utilization: each component, including AWGs,
TWCs, and internal interconnection links, should be fully utilized
when all input and output wavelength channels are busy. In respect
to these non-blocking criteria, a comparison between previous works
and our approach on the design of AWG-based switching networks is
described below.

\subsection{Previous Works}
\begin{figure}[t]
\centering
\includegraphics[width=0.3\textwidth]{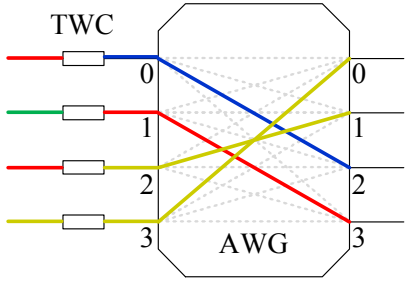}
\caption{A $4\times4$ AWG-based crossbar.}\label{f_awgcrossbar}
\end{figure}

A straightforward construction approach of an $N\times N$ AWG-based
switching network is to install a TWC at each input of an $N\times
N$ AWG \cite{Blumenthal:JSTQE,Cheyns:PNC2003,Pan:JLT2005}. The
network is logically equivalent to an $N\times N$ crossbar and
referred to as an \emph{AWG-based crossbar}. Despite the fact that
this network is strictly non-blocking (SNB), it is not scalable
because the size of the AWG, the conversion range of the TWCs, and
the wavelength granularity are all equal to $N$. When $N$ is large,
the wavelength granularity of the switching network may be greater
than the number of the wavelengths carried by each input or output
fiber, and the network suffers from severe crosstalk in AWG and the
high cost of the TWC conversion range. Furthermore, the utilization
of the AWG is only $1/N$, because the AWG-based crossbar can only
establish $N$ connections at most while the AWG can provide $N^2$
interconnections, or wavelength channels, between the inputs and
outputs \cite{Ye:JLT2012} at the same time. An underutilized
$4\times4$ AWG-based crossbar is depicted in Fig.
\ref{f_awgcrossbar}, the dotted gray lines represent idle wavelength
channels in the AWG while all inputs are busy.

Several methods have been proposed to suppress the crosstalk of
AWG-based switches
\cite{Bianco:TON,Lucerna:HPSR2011,Maier:SPIE2000}. A
crosstalk-preventing scheduling algorithm was introduced in
\cite{Bianco:TON} that prohibits using the same wavelength at more
than 15 inputs in an AWG. Despite suppressing coherent crosstalk,
this algorithm significantly reduces the number of wavelength
channels in an AWG, and requires additional computation complexity
to find admissible states. The method introduced in
\cite{Lucerna:HPSR2011} can achieve a zero coherent crosstalk (ZCC)
switch by using a restricted subset of AWG inputs. For example,
according to this method, a $32\times 32$ AWG is used to construct a
$8\times8$ ZCC switch, in which the AWG is severely underutilized
yet the required conversion range of the TWCs is greater than 8. The
scheme proposed in \cite{Maier:SPIE2000} exploits the wavelengths in
multiple free spectrum ranges (FSRs) to prevent using the same
wavelength at multiple inputs. This method is not practical because
the AWG performs poorly outside the main FSR
\cite{Kamei:JLT2009,Liu:EL1999}, and the wavelength granularity of
the switching network and the conversion range of the TWCs could be
several times larger than the port count $N$.

A single-stage network, sandwiching an array of $W\times W$
AWG-based crossbars between $F$ input fibers and $F$ output fibers
by a set of interconnection links, was proposed in
\cite{Pattavina:INFOCOM2006,Ramamirtham:INFOCOM2002}, where each
fiber carries $W$ wavelengths. This network is typically internally
blocking. It is not scalable because the port count of the AWGs, the
conversion range of the TWCs, and the wavelength granularity are all
equal to $W$, the number of wavelengths per fiber, which could be
large in practice. Moreover, the maximum utilization of each AWG is
only $1/W$ even when all input and output channels are fully loaded.

The multi-stage AWG-based switching networks were considered in
\cite{Cheyns:CM2004,Ngo:TON2006,Zanzottera:HPSR2006}, in which a set
of AWG-based crossbars are interconnected to form a three-stage
rearrangeably non-blocking Clos type network \cite{Clos:BSTJ1953}.
This switch architecture scales down the size of the AWGs, the
conversion range of the TWCs, and the wavelength granularity, but
does not fully utilize the wavelength routing properties of the
AWGs. In fact, each internal link in this network only carries one
wavelength and the routing is performed in the same manner as that
in a pure space-division network. As a result, this design exhibits
two drawbacks. First, the utilization of the AWGs remains low
because each modular switch is an AWG-based crossbar. Second, the
physical interconnection complexity is high because the number of
links between two adjacent stages is equal to the number of input
wavelength channels $N$.

Alternate AWG-based multi-stage networks were proposed in
\cite{Bregni:JSAC2003,Maier:CM2004,Zhong:JLT1996,Leonardus:EU2007},
in which each input of the AWGs is equipped with an array of TWCs,
such that each internal link carries a set of wavelength channels.
These switch architectures were designed to fully utilize the
wavelength routing properties of the AWGs and significantly reduce
the physical interconnection complexity. Though the switching
networks proposed in \cite{Zhong:JLT1996,Leonardus:EU2007} are
non-blocking, the relevant routing algorithms were not presented.
Moreover, the size of the AWGs, the conversion range of the TWCs,
and the wavelength granularity in these networks are still
increasing with respect to the number of the input wavelength
channels $N$.

\subsection{Our Approach and Contribution}
This paper illustrates the recursive construction principle of an
AWG-based non-blocking Clos network, by taking the network
scalability and optimal utilization into consideration. We first
apply the technique of modular decomposition of AWGs developed in
\cite{Ye:JLT2012} to the three-stage AWG-based networks. The initial
goal of the design is to modularize the AWGs and partition the
wavelength set while preserving the wavelength routing properties of
AWGs. Next, we introduce the concept of wavelength independence
based on the fictitious boundaries that lie in the middle of TWCs.
We show that the entire WDM switch network can be divided into
wavelength autonomous cells. Consequently, both the wavelength
granularity of the network and the conversion range of the TWCs can
be substantially reduced by reusing the similar, yet smaller, sets
of wavelengths. As for non-blocking route assignments, we show that
the rearrangeability of these AWG-based three-stage networks is the
same as that of classical Clos networks. In sum, the contribution of
this paper is to show that our proposed three-stage AWG-based Clos
network can accomplish the following desirable objectives:
\begin{itemize}
[\IEEEsetlabelwidth{NB}]
\item[NB1] \emph{The rearrangeably non-blocking (RNB) three-stage network with the minimum number of central modules
can achieve 100\% utilization when all inputs and outputs are fully
loaded},
\item[NB2] \emph{Modularize AWGs},
\item[NB3] \emph{Scale down the conversion range of TWCs}, and
\item[NB4] \emph{Reuse the same wavelength set in the recursive construction of the network to reduce the wavelength
granularity}.
\end{itemize}
The objective NB1 is consistent with the RNB condition of the
traditional Clos networks, while NB2$\sim$NB4 are additional
requirements for WDM-based Clos networks.

\begin{figure}[t]
\centering
\includegraphics[width=0.49\textwidth]{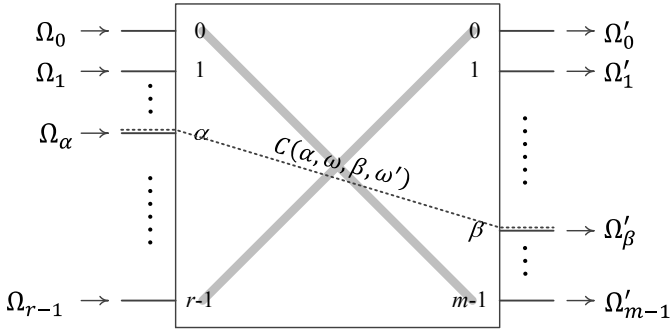}
\caption{A call in a WDM switch.}\label{f_wdm-swtich}
\end{figure}

The rest of this paper is organized as follows. In Section
\ref{preliminary}, we briefly introduce the function of the AWGs and
TWCs, and show how a WDM switching network can be constructed from
the combination of AWGs and TWCs. In Section \ref{RNB_design}, we
construct a three-stage AWG-based Clos network, based on the
non-blocking and contention-free principle of WDM switches. We
present the routing algorithm of this network, and show that the
network is fully utilized if all input and output wavelength
channels are busy. Section \ref{recurConstruction} proposes a
recursive construction scheme to achieve the scalability of
AWG-based Clos networks. We show that this scheme scales down the
AWGs, and substantially reduces wavelength granularity of the
network and, thus, the conversion range of the TWCs. Section
\ref{conclusion} concludes this paper with a comparison of our
results to previous results.

\section{Preliminaries}\label{preliminary}
In this paper, we study the wavelength-based communication model
\cite{Pattavina:INFOCOM2006}. The WDM switch under consideration is
illustrated in Fig. \ref{f_wdm-swtich}. This switch has $r$ input
ports and $m$ output ports, and each input port and each output port
carries $m$ wavelengths and $r$ wavelengths, respectively. Thus, the
dimension of this WDM switching network is $N\times N$, where $N=rm$
is the number of input (output) wavelength channels. The input ports
are numbered by $0,1,\cdots,r-1$ from the top to the bottom. The set
of wavelengths carried by the input port $\alpha$ is denoted by
$\Omega_\alpha$, for $\alpha=0,1,\cdots,r-1$. Similarly, the output
ports are labeled by $0,1,\cdots,m-1$, and the set of wavelengths
associated with output port $\beta$ is denoted as $\Omega_\beta'$
for $\beta=0,1,\cdots,m-1$. Furthermore, without loss of generality,
we assume that input wavelength sets are all the same, (i.e.,
$\Omega_\alpha=\Omega$), and the output wavelength sets are all the
same (i.e., $\Omega_\beta'=\Omega'$).

Let $I(\alpha,\omega)$ denote the input channel at the input port
$\alpha$ carried by the wavelength $\omega\in\Omega$, and
$O(\beta,\omega')$ be the output channel at the output port $\beta$
carried by the wavelength $\omega'\in\Omega'$. As illustrated by the
dotted line in Fig. \ref{f_wdm-swtich}, a call
$C(\alpha,\omega,\beta,\omega')$ in the WDM switch is defined as a
connection between the input channel $I(\alpha,\omega)$ and the
output channel $O(\beta,\omega')$. This paper focuses on the
AWG-based WDM switches with rearrangeably non-blocking (RNB)
properties, meaning that a call can always be established between an
idle input channel $I(\alpha,\omega)$ and an idle output channel
$O(\beta,\omega')$ with possible rearrangements of existing
connections. To facilitate our discussion, we first describe the
functions of AWG and TWC in a WDM switch.

\subsection{AWG}
\begin{figure}[t]
\centering
\includegraphics[width=0.49\textwidth]{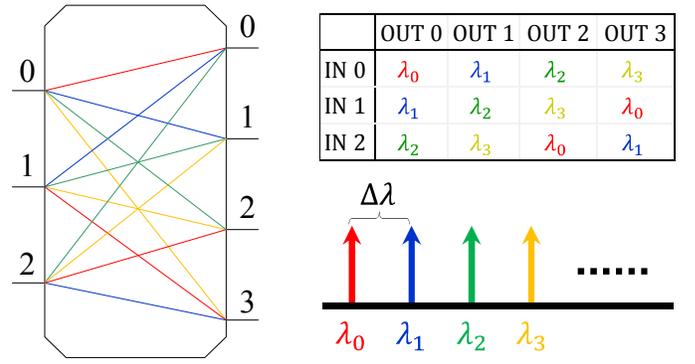}
\caption{A $3\times4$ AWG and its routing table.}\label{f_awg}
\end{figure}

An $r\times m$ AWG is associated with a set of equally spaced
wavelengths $\Lambda=\{\lambda_0,\cdots,\lambda_{|\Lambda|-1}\}$ in
its principal FSR, where $|\Lambda|=\max\{r,m\}$. The AWG has a
cyclic wavelength routing property: the signal carried by the
wavelength $\lambda_i\in\Lambda$ at input $j$ will be forwarded to
output $k$, if
\begin{equation}\label{equ_awg_routing1}
\centering [i-j]_{|\Lambda|}=k,
\end{equation}
or equivalently,
\begin{equation}\label{equ_awg_routing2}
\centering [j+k]_{|\Lambda|}=i,
\end{equation}
where $[x]_y\stackrel{\Delta}{=}(x~\textrm{mod}~y)$. From
(\ref{equ_awg_routing2}), it is easy to see that each input port of
the AWG is associated with a wavelength subset of $\Lambda$. Let
$\Lambda_a$ be the wavelength subset associated with the $a$th input
port, where $a=0,1,\cdots,r-1$. According to
(\ref{equ_awg_routing2}), we have
\[\Lambda_a=\{\lambda_a,\lambda_{[a+1]_{|\Lambda|}},\cdots,\lambda_{[a+m-1]_{|\Lambda|}}\}.\]
Similarly, let $\Lambda_b'$ be the wavelength subset associated with
the $b$th output port, where $b=0,1,\cdots,m-1$. We have
\[\Lambda_b'=\{\lambda_b,\lambda_{[b+1]_{|\Lambda|}},\cdots,\lambda_{[b+r-1]_{|\Lambda|}}\}.\]
It is clear that
$\bigcup_{a=0}^{r-1}\Lambda_a=\bigcup_{b=0}^{m-1}\Lambda_b'
=\Lambda$. We have shown in \cite{Ye:JLT2012} that such wavelength
assignment of the AWGs is contention-free. As an example, the
wavelength assignments of the $3\times4$ AWG shown in Fig.
\ref{f_awg} are tabulated in a contention-free routing table, in
which the output port 2 is associated with the wavelength subset
$\Lambda_2'=\{\lambda_2,\lambda_3,\lambda_0\}$. For a symmetric AWG,
the routing table is a Latin square.

The $r\times m$ AWG provides a set of $rm$ fixed interconnection
channels between its input ports and output ports via a group of
$|\Lambda|$ wavelengths \cite{Ye:JLT2012}. For example, the 12
interconnection channels between the inputs and the outputs of a
$3\times4$ AWG are shown in Fig. \ref{f_awg}. Larger AWGs can
provide much richer interconnection channels; however, the results
in \cite{Gaudino:ICC2008} show that the AWG with a large port count
suffers serious coherent crosstalk, which eventually imposes a
limitation on the scalability of AWG-based optical switches
\cite{Bianco:TON}.

\subsection{TWC-Module}
\begin{figure}[t]
\centering
\includegraphics[width=0.24\textwidth]{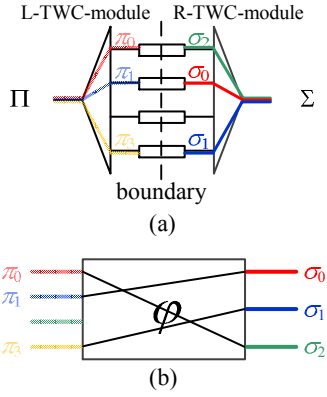}
\caption{Illustration of a TWC-module: (a) $4\times3$ TWC-module,
and (b) space representation.}\label{f_twc}
\end{figure}

A set of $n$ TWCs sandwiched by a $1\times n$ demultiplexer (DeMux)
and an $s\times1$ optical combiner (Mux) composes an $n\times s$
TWC-module, which can convert the wavelengths in the domain set
$\Pi=\{\pi_0,\cdots,\pi_{n-1}\}$ at the input of the DeMux to the
wavelengths in the range set
$\Sigma=\{\sigma_0,\cdots,\sigma_{s-1}\}$ at the output of the Mux.
Specifically, the function of a TWC-module is to perform a
wavelength conversion mapping $\phi:\Pi\rightarrow\Sigma$. The
conversion capability of each TWC is characterized by the size of
range set $\Sigma$, or $s$, which is referred to as the
\emph{conversion range}. An example is the $4\times3$ TWC-module
shown in Fig. \ref{f_twc}(a), where
$\Pi=\{\pi_0,\pi_1,\pi_2,\pi_3\}$ and
$\Sigma=\{\sigma_0,\sigma_1,\sigma_2\}$. In this TWC-module, the
conversion range of each TWC is 3, and the wavelengths $\pi_0$,
$\pi_1$, and $\pi_3$ are converted to $\sigma_2$, $\sigma_0$, and
$\sigma_1$, respectively. The wavelength conversion mapping of a
TWC-module can also be viewed as a permutation pattern in a
crossbar, as shown in Fig. \ref{f_twc}(b), where each port
represents a wavelength. A TWC with a larger conversion range is
more powerful but at a much higher cost
\cite{Pattavina:INFOCOM2006}.

In this paper, we focus on the recursive construction of modular WDM
switch architectures, in which the wavelength sets associated with
different modular switches must be independent of each other.
Therefore, it is sometimes necessary to separate the domain set
$\Pi$ and the range set $\Sigma$ of a TWC-module. Logically, a
TWC-module can be divided by a fictitious boundary into two parts,
the L-TWC-module and the R-TWC-module, as indicated by the dashed
line displayed in Fig. \ref{f_twc}(a), while the conversion mapping
$\phi:\Pi\rightarrow\Sigma$ is performed at the boundary between the
L-TWC-module and the R-TWC-module.

\subsection{Two-Stage AWG-Based Network}
\begin{figure}[t]
\centering
\includegraphics[width=0.49\textwidth]{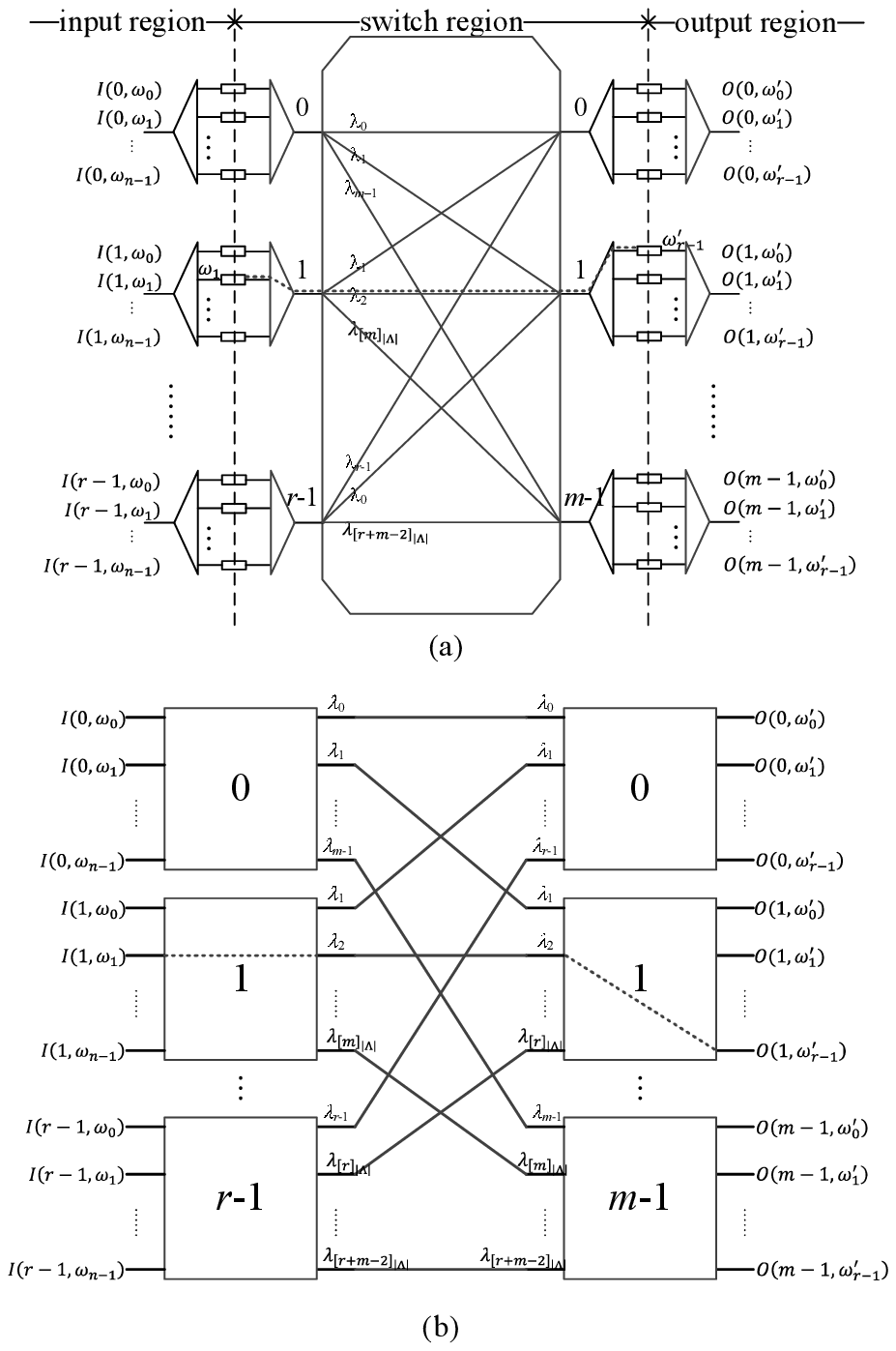}
\caption{A two-stage AWG-based network: (a) $\mathcal{T}(n,r,m)$ and
(b) space representation.}\label{f_banyan}
\end{figure}

In this part, we consider a WDM switch that is composed of a single
$r\times m$ AWG and a set of TWC-modules. In this switch, each input
port is equipped with an $n\times m$ input TWC-module, and each
output port is equipped with an $r\times r$ output TWC-module. This
WDM switch is associated with a wavelength set
$\Lambda=\{\lambda_0,\cdots,\lambda_{|\Lambda|-1}\}$, and thus its
wavelength granularity is $|\Lambda|=\max\{r,m\}$. The labeling of
$r$ input TWC-modules and $m$ output TWC-modules is displayed in
Fig. \ref{f_banyan}(a). This network is referred to as a
\emph{two-stage AWG-based network} and denoted by
$\mathcal{T}(n,r,m)$.

The task of the TWC-modules is to perform wavelength conversions.
Let $\Omega$ and $\Omega'$ be the wavelength sets carried by input
and output channels, respectively. In the input TWC-module $\alpha$,
the input channels carried by the wavelengths in the set $\Omega$
are fully demultiplexed by the L-TWC-module, and their wavelengths
are converted to the wavelengths in the set $\Lambda_\alpha$, i.e.,
performing the wavelength mapping
$\phi_\alpha:\Omega\rightarrow\Lambda_\alpha$, where
$|\Lambda_\alpha|=m$. Similarly, in the output TWC-module $\beta$,
the wavelength mapping
$\phi_\beta':\Lambda_\beta'\rightarrow\Omega'$ is carried out at the
boundary. The wavelengths in the set $\Omega'$ are multiplexed
together by the R-TWC-module and fed to the output port. Therefore,
through wavelength conversions, the input and output TWC-modules
form two fictitious boundaries that separate the network into three
parts: input wavelength region, switch wavelength region, and output
wavelength region, which are associated with $\Omega$, $\Lambda$,
and $\Omega'$, respectively.

According to the wavelength routing property of AWGs defined by
(\ref{equ_awg_routing2}), there is a unique connection between each
input TWC-module and each output TWC-module in this switch. A call
request from the input channel $I(\alpha,\omega)$ to the output
channel $O(\beta,\omega')$, denoted as
$C(\alpha,\omega,\beta,\omega')$, can be established according to
(\ref{equ_awg_routing2}) by the wavelength $\lambda_x$ in the AWG,
where $x=[\alpha+\beta]_{|\Lambda|}$. The input TWC-module $\alpha$
translates the input wavelength $\omega\in\Omega$ to $\lambda_x$,
and the output TWC-module $\beta$ converts $\lambda_x$ to the output
wavelength $\omega'\in\Omega'$. The route and wavelength assignment
(RWA) for the call $C(\alpha,\omega,\beta,\omega')$ can be expressed
as follows:
\[I(\alpha,\omega)\rightarrow \underset{\textrm{stage
1}}{\underbrace{\alpha}}
\xrightarrow{\lambda_{[\alpha+\beta]_{|\Lambda|}}}\underset{\textrm{stage
2}}{\underbrace{\beta}}\rightarrow O(\beta,\omega'),\] which is also
illustrated in Fig. \ref{f_banyan}(a). On the other hand, since only
one wavelength can be assigned to a paired input TWC-module and
output TWC-module, this two-stage AWG-based network is internally
blocking. Two input channels that share the same input TWC-module
cannot be switched to the same output TWC-module at the same time.
The reason that the two-stage AWG network is blocking is logically
equivalent to that of a Banyan network \cite{Ttlee:book2011}. The
analogy between the two-stage AWG network and the Banyan network is
shown in Fig. \ref{f_banyan}(b), which demonstrates the following
one-to-one correspondences between these two networks:
\begin{itemize}
\item[(1)] a crossbar module and a TWC-module,
\item[(2)] an input/output port and an input/output wavelength,
\item[(3)] a link between the crossbars at two adjacent stages and a wavelength route in the AWG.
\end{itemize}
Therefore, the two-stage AWG-based network is also referred to as
\emph{AWG-based Banyan network} in this paper. In the next section,
we show that a non-blocking WDM switch can be realized by a
three-stage AWG-based Clos network, which is the cascaded
combination of two AWG-based Banyan networks.

\section{Rearrangeably Non-blocking AWG-Based Three-Stage Networks}\label{RNB_design}
In this section, we explore the recursive construction of AWG-based
three-stage networks. Since connections carried by different
wavelengths can be multiplexed into a single link, the non-blocking
condition of a WDM switch is different from that of a space-division
network. Specifically, the following fundamental constraint is
imposed on all connections in a multi-stage non-blocking WDM switch
network:

\begin{figure*}[t]
\centering
\includegraphics[width=0.72\textwidth]{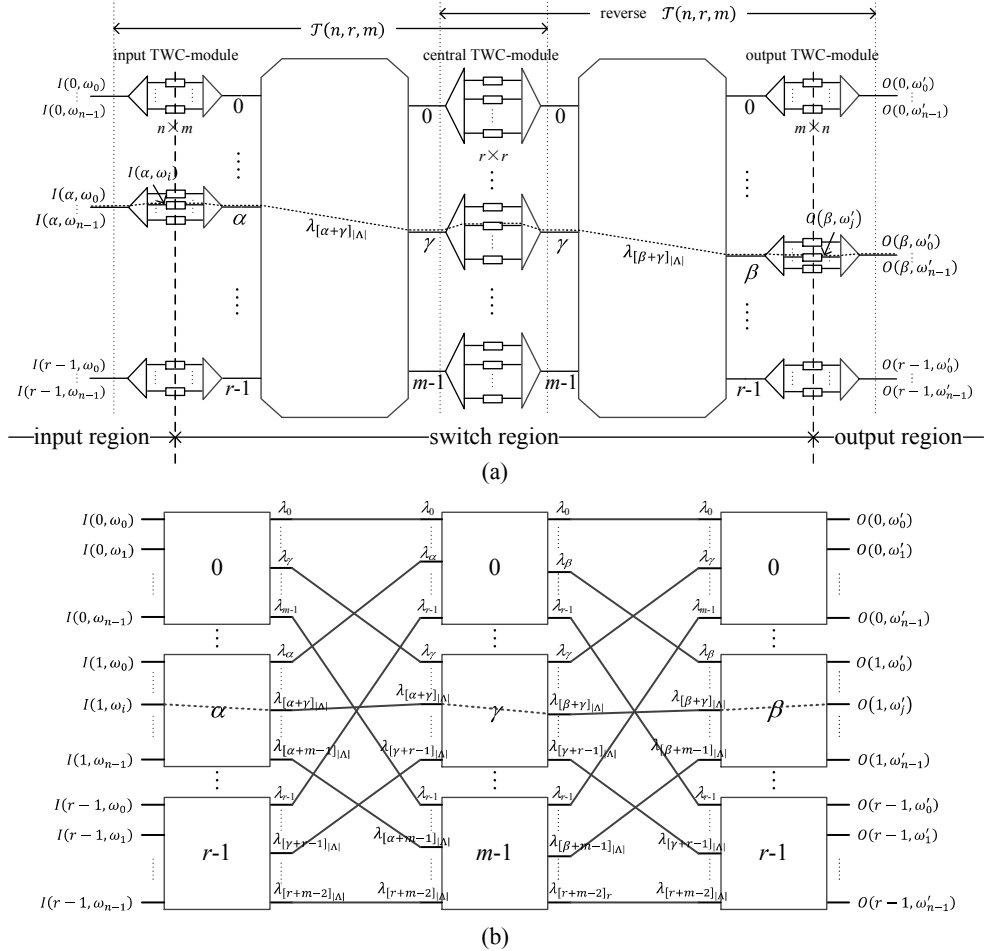}
\caption{AWG-based three-stage network: (a) $\mathcal{S}_A(n,r,m)$
and (b) space representation.}\label{f_clos}
\end{figure*}

\textbf{Non-blocking and Contention-free Principle of WDM Switches:}
\emph{If two connections share a common link, then they must be
carried by different wavelengths, or equivalently, calls carried by
the same wavelength must be connected by separated links.}

\begin{figure*}[t]
\centering
\includegraphics[width=0.72\textwidth]{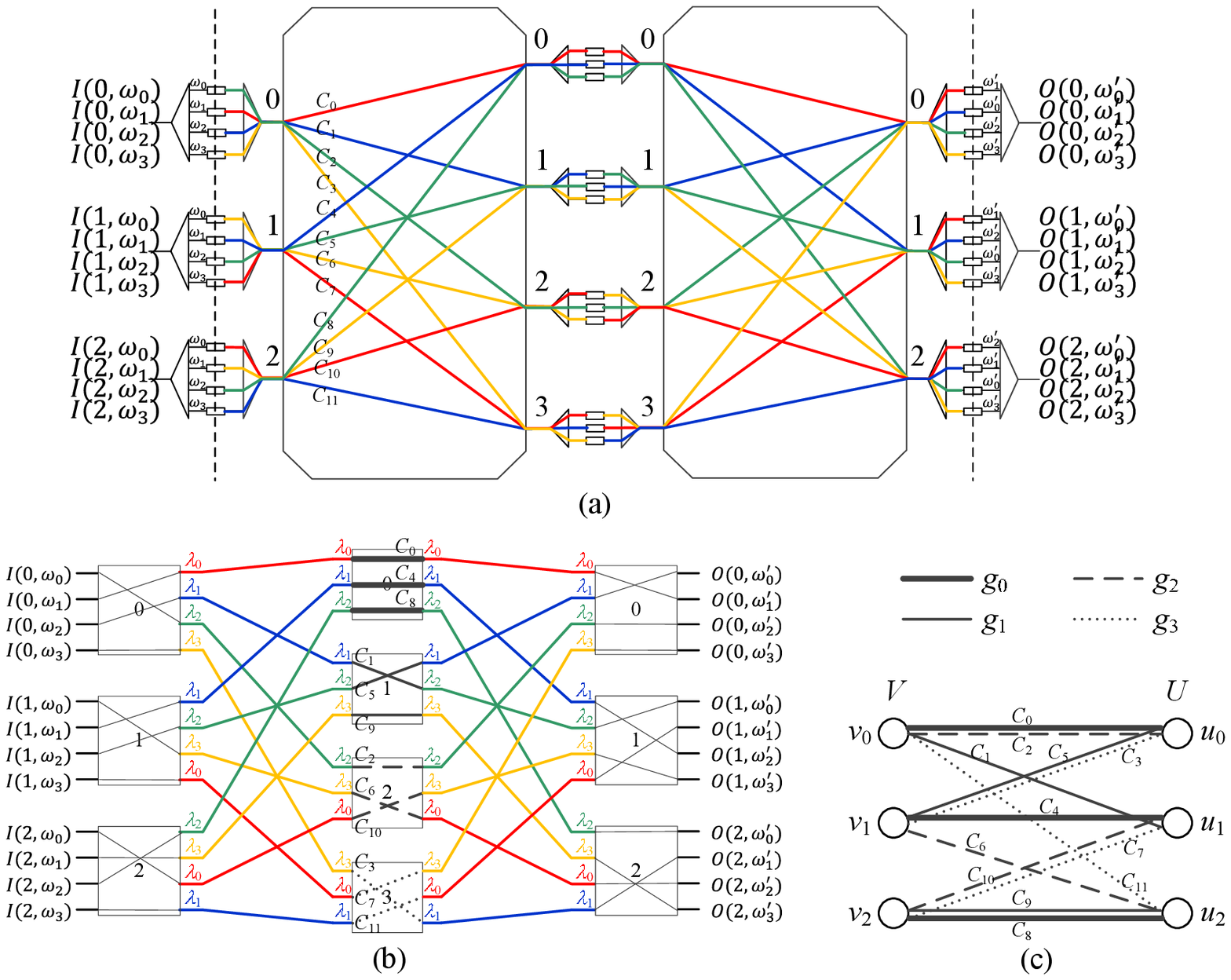}
\caption{An AWG-based three-stage network: (a)
$\mathcal{S}_A(4,3,4)$, (b) space representation, and (c) conflict
graph.}\label{f_clos-example}
\end{figure*}

As Section \ref{preliminary} illustrates, the AWG-based Banyan
network is internally blocking because there is only one link
between each input TWC-module and each output TWC-module. In a
three-stage network, similar to the space-division Clos network, the
number of alternative paths between a pair of input and output
TWC-modules can be increased by the cascaded combination of two
AWG-based Banyan networks. Furthermore, we show in this section that
the non-blocking three-stage AWG-based network with the minimum
number of central modules can accomplish the desirable objectives
NB1$\sim$NB4 mentioned in Section \ref{intro}.

The $N\times N$ three-stage AWG-based network under consideration,
as shown in Fig. \ref{f_clos}(a), is a combination of a
$\mathcal{T}(n,r,m)$ network and a reverse $\mathcal{T}(n,r,m)$
network, denoted as $\mathcal{S}_A (n,r,m)$, where $N=rn$ is the
number of wavelength channels and $m$ is the number of central
TWC-modules. The wavelength granularity of the network
$\mathcal{S}_A (n,r,m)$ is $|\Lambda|$, which is the same as that of
the $\mathcal{T}(n,r,m)$. The two fictitious boundaries in the
middle of the input and output TWC-modules are represented by the
two dashed lines in Fig. \ref{f_clos}(a). They divide the entire
network into three wavelength independent regions, corresponding to
the external input wavelength set $\Omega$, the switch wavelength
set $\Lambda$, and the external output wavelength set $\Omega'$,
respectively.

In the network $\mathcal{S}_A (n,r,m)$, an input channel
$I(\alpha,\omega_i)$ on the input TWC-module $\alpha$ and an output
channel $O(\beta,\omega_j')$ on the output TWC-module $\beta$ can be
connected via a central TWC-module $\gamma$, for
$\gamma=0,1,\cdots,m-1$. According to the wavelength routing
property of the AWG, the central TWC-module $\gamma$ is connected to
the input TWC-module $\alpha$ by wavelength $\lambda_x$, and the
output TWC-module $\beta$ by wavelength $\lambda_y$, where the
indices are given by
\begin{equation}\label{equ_awgrouting_clos1}
\centering x=[\alpha+\gamma]_{|\Lambda|},
\end{equation}
and
\begin{equation}\label{equ_awgrouting_clos2}
\centering y=[\beta+\gamma]_{|\Lambda|}.
\end{equation}
The wavelength $\lambda_x$ is converted to $\lambda_y$ in the middle
by the central TWC-module $\gamma$. The connection of the call
$C(\alpha,\omega_i,\beta,\omega_j')$, as shown in Fig. \ref{f_clos},
is given by: \setlength{\arraycolsep}{0.1em}
\begin{eqnarray}\centering
I(\alpha,\omega_i) & \rightarrow & \underset{\textrm{stage
1}}{\underbrace{\alpha}}\xrightarrow{\lambda_{[\alpha+\gamma]_{|\Lambda|}}}
\underset{\textrm{stage
2}}{\underbrace{\gamma}}\xrightarrow{\lambda_{[\gamma+\beta]_{|\Lambda|}}}\underset{\textrm{stage
3}}{\underbrace{\beta}}\nonumber\\
& \rightarrow & O(\beta,\omega_j'),\nonumber
\end{eqnarray}
which contains the following sequence of wavelength conversions:
\begin{itemize}
\item input TWC-module $\alpha:\omega_i\rightarrow\lambda_x=\lambda_{[\alpha+\gamma]_{|\Lambda|}}$,
\item central TWC-module $\gamma:\lambda_x\rightarrow\lambda_y$,
\item output TWC-module $\beta:\lambda_y=\lambda_{[\beta+\gamma]_{|\Lambda|}}\rightarrow\omega_j'$.
\end{itemize}
Furthermore, if we replace every TWC with a crossbar switch, the
$\mathcal{S}_A (n,r,m)$ is logically equivalent to a symmetric
three-stage Clos network \cite{Ttlee:book2011} with the inter-stage
connections specified by (\ref{equ_awgrouting_clos1}) and
(\ref{equ_awgrouting_clos2}) as shown Fig. \ref{f_clos}(b).

Similar to the non-blocking condition of Clos networks, two calls
$C_1(\alpha_1,\omega_{i_1},\beta_1,\omega_{j_1}')$ and
$C_2(\alpha_2,\omega_{i_2},\beta_2,\omega_{j_2}')$ in the
$\mathcal{S}_A (n,r,m)$ cannot share the same central TWC-module if
they are from the same input TWC-module, i.e.,
$\alpha_1=\alpha_2=\alpha$, or if they are destined for the same
output TWC-module, i.e., $\beta_1=\beta_2=\beta$. It follows that
the contentions among calls in the $\mathcal{S}_A (n,r,m)$ can be
modeled as a bipartite graph $G(V\cup U,E)$, in which the vertex
$v_\alpha\in V$ and the vertex $u_\beta\in U$ denote the input
TWC-module $\alpha$ and the output TWC-module $\beta$, respectively,
and a call $C(\alpha,\omega,\beta,\omega')$ is represented by an
edge $e_{\alpha\beta}\in E$ that connects $v_\alpha$ and $u_\beta$.
The bipartite graph $G(V\cup U,E)$ is $n$-regular with $2r$ vertices
and $rn$ edges if all input and output wavelength channels in the
network $\mathcal{S}_A (n,r,m)$ are busy.

Let $\Gamma$ be a set of colors, such that each color
$g_\gamma\in\Gamma$ corresponds to a central TWC-module $\gamma$.
The fact that two calls in conflict cannot share the same central
TWC-module corresponds to the condition that two edges terminated on
the same vertex in $G(V\cup U,E)$ cannot be colored with the same
color. Thus, the RWA problem of $\mathcal{S}_A (n,r,m)$ can be
solved by the edge coloring of the bipartite graph $G(V\cup U,E)$.
As an example, we consider the calls in the network $\mathcal{S}_A
(4,3,4)$ shown in Fig. \ref{f_clos-example}(a), in which the
connections were established by the 4-edge coloring of the bipartite
graph displayed in Fig. \ref{f_clos-example}(c). In this example,
the four central TWC-modules, namely 0, 1, 2, and 3, are represented
by color $g_0$, $g_1$, $g_2$, and $g_3\in\Gamma$, respectively.
According to the edge coloring shown in Fig.
\ref{f_clos-example}(c), the corresponding route assignments, as
illustrated in Fig. \ref{f_clos-example}(b), are given in Table
\ref{t_RWA}. Once the central module assigned to each call is fixed,
the wavelengths that carry out the connections can be immediately
determined by (\ref{equ_awgrouting_clos1}) and
(\ref{equ_awgrouting_clos2}), which are also shown in Table
\ref{t_RWA}.

Many efficient algorithms for edge coloring of bipartite graphs are
available. For example, using the edge coloring algorithms reported
in \cite{Cole:C2001} and \cite{Ttlee:IJCM2013} to color the
bipartite graph $G(V\cup U,E)$ of the network $\mathcal{S}_A
(n,r,m)$, the time complexities are on the order of $O(nr\log n)$
and $O(nr\log 2r)$, respectively. If the switch contains $n$ central
TWC-modules and $n$ is an integral power of 2, then the time
complexity can be further reduced to $O(\log nr\times\log n)$ by
using the algorithm described in \cite{Ttlee:TCOM2002}. Heuristic
algorithms for bipartite matching, such as iSLIP
\cite{McKeown:TON1999} and DRRM \cite{Chao:book2001}, are also
available for online implementation of packet switching systems.

In graph theory, the chromatic index $\chi(G)$ of a graph $G$ is the
minimum number $k$ for which $G$ is $k$-edge colorable. From Hall's
theorem \cite{Hall:JLMS1936}, it is easy to prove that the chromatic
index $\chi(G)$ of a bipartite graph $G$ is equal to the maximum
vertex degree $\Delta$ of $G$. Since the bipartite $G(V\cup U,E)$ of
the network $\mathcal{S}_A (n,r,m)$ has a maximum vertex degree
$\Delta=n$, we immediately obtain the following condition on RNB
networks.

\begin{theorem}\label{theorem_SA}
The network $\mathcal{S}_A (n,r,m)$ is RNB if and only if $m\ge n$.
\end{theorem}

In classic Clos networks, the proof of the above theorem was
originally given by Slepian \cite{Slepian:U1952} and Duguid
\cite{Duguid:T1959}. Their proof of the theorem is actually a
rearrangement algorithm for blocked calls. The same algorithm can be
applied to the network $\mathcal{S}_A (n,r,m)$ in case blocking
occurred. According to the above theorem, the minimum number of
central TWC modules required by a rearrangeably non-blocking network
$\mathcal{S}_A (n,r,m)$ is $m=n$, in which case the network
$\mathcal{S}_A (n,r,n)$ has exactly $N=rn$ TWCs at each stage and
$N=rn$ wavelength channels within each AWG. Thus, similar to the
symmetric rearrangeably non-blocking (RNB) Clos networks with
minimum number of central modules, the optimal network
$\mathcal{S}_A (n,r,n)$ can also achieve 100\% utilization when all
inputs and outputs are fully loaded with $N=rn$ calls.

\begin{table}[t]
\centering \caption{Route and wavelength assignment (RWA) for the
calls in Fig. \ref{f_clos-example}}
\begin{tabular}{||c|c|c|c||}
\hline  & Input & Central TWC-module & output \\
\rb{call} &  TWC-module & (wavelength conversion) & TWC-module \\
\hline\hline $C_0$ &  0 & 0 ($\textcolor{red}{\lambda_0}\rightarrow\textcolor{red}{\lambda_0}$) & 0 \\
\hline $C_4$ & 1 & 0 ($\textcolor{blue}{\lambda_1}\rightarrow\textcolor{blue}{\lambda_1}$) & 1 \\
\hline $C_8$ & 2 & 0 ($\textcolor{ForestGreen}{\lambda_2}\rightarrow\textcolor{ForestGreen}{\lambda_2}$) & 2 \\
\hline\hline $C_1$ & 0 & 1 ($\textcolor{blue}{\lambda_1}\rightarrow\textcolor{ForestGreen}{\lambda_2}$) & 1 \\
\hline $C_5$ & 1 & 1 ($\textcolor{ForestGreen}{\lambda_2}\rightarrow\textcolor{blue}{\lambda_1}$) & 0 \\
\hline $C_9$ & 2 & 1 ($\textcolor{YellowOrange}{\lambda_3}\rightarrow\textcolor{YellowOrange}{\lambda_3}$) & 2 \\
\hline\hline $C_2$ & 0 & 2 ($\textcolor{ForestGreen}{\lambda_2}\rightarrow\textcolor{ForestGreen}{\lambda_2}$) & 1 \\
\hline $C_6$ & 1 & 2 ($\textcolor{YellowOrange}{\lambda_3}\rightarrow\textcolor{red}{\lambda_0}$) & 0 \\
\hline $C_{10}$ & 2 & 2 ($\textcolor{red}{\lambda_0}\rightarrow\textcolor{YellowOrange}{\lambda_3}$) & 2 \\
\hline\hline $C_3$ & 0 & 3 ($\textcolor{YellowOrange}{\lambda_3}\rightarrow\textcolor{blue}{\lambda_1}$) & 0 \\
\hline $C_7$ & 1 & 3 ($\textcolor{red}{\lambda_0}\rightarrow\textcolor{YellowOrange}{\lambda_3}$) & 1 \\
\hline $C_{11}$ & 2 & 3 ($\textcolor{blue}{\lambda_1}\rightarrow\textcolor{red}{\lambda_0}$) & 2 \\
\hline
\end{tabular}\label{t_RWA}
\end{table}

The major goal of the classic three-stage Clos networks is to scale
crossbar modules by using a recursive decomposition scheme. However,
the AWGs and the conversion range of central TWC-modules in the
network $\mathcal{S}_A (n,r,n)$  are not scalable. It also emerged
as the most intractable issue of similar AWG-based three-stage
networks proposed in \cite{Zhong:JLT1996} and
\cite{Leonardus:EU2007}. The complete recursive scheme to achieve
the scalability of AWG-based Clos networks is described in the next
section.

\begin{figure*}[t]
\centering
\includegraphics[width=0.6\textwidth]{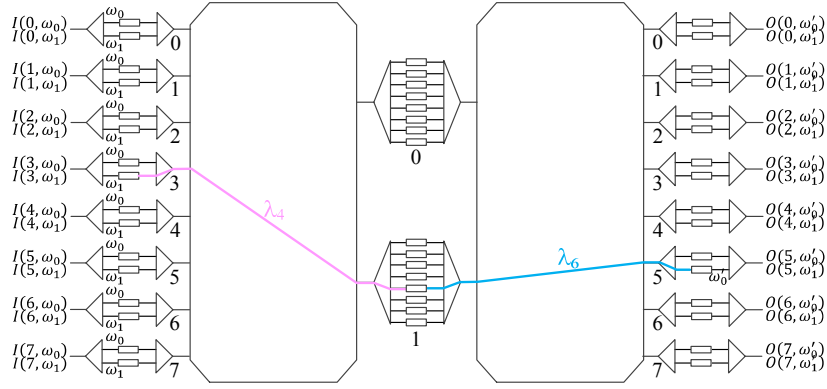}
\caption{An AWG-based three-stage network
$\mathcal{S}_A(2,2^{4-1},2)$.}\label{f_sa-example}
\end{figure*}

\section{Recursive Construction: AWG Modularization and Wavelength Reuse}\label{recurConstruction}
To achieve our objective NB2$\sim$NB4, we focus our discussion on
the modularization of AWGs and the rule to reduce the conversion
range of the central TWCs. Without loss of generality, we consider
the recursive construction of the network $\mathcal{S}_A (n,r,n)$
with parameter $r=n^{d-1}$. An example network is illustrated in
Fig. \ref{f_sa-example}, where $n=2$ and $d=4$. This $n^d\times n^d$
three-stage switch network consists of an $n^{d-1}\times n$ input
AWG and an $n\times n^{d-1}$ output AWG with the wavelength set
$\Lambda=\{\lambda_0,\lambda_1,\cdots,\lambda_{n^{d-1}-1}\}$ in the
switch region. Thus, the wavelength granularity of the network
$\mathcal{S}_A (n,n^{d-1},n)$ is $n^{d-1}$. Also, the conversion
range of each input or output TWC is $n$, while that of central TWCs
is $n^{d-1}$.

\begin{figure}[t]
\centering
\includegraphics[width=0.32\textwidth]{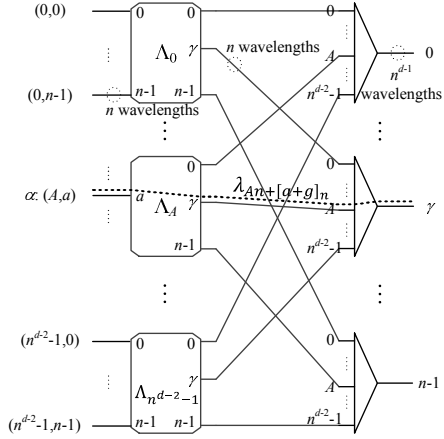}
\caption{Two-stage AWG network
$\mathcal{N}_I(n^{d-2},n)$.}\label{f_awg-network}
\end{figure}

We first consider the decomposition of an $n^{d-1}\times n$ AWG into
a two-stage network, denoted as $\mathcal{N}_I (n^{d-2},n)$. As
shown in Fig. \ref{f_awg-network}, the network consists of $n^{d-2}$
$n\times n$ AWGs in the first stage and $n$ $n^{d-2}\times1$ Muxes
in the second stage. In the two-stage network, the input $\alpha$ is
labeled by the two-tuple $(A,a)$, where $A=\lfloor\alpha/n\rfloor$
and $a=[\alpha]_n$, to indicate that it is the $a$th input of the
$A$th AWG module. Herein, $\lfloor Y\rfloor$ denotes the largest
integer smaller than $Y$. The set of wavelengths associated with AWG
$A$ is a subset $\Lambda_A\subseteq\Lambda$, defined by
\begin{equation}\label{equ_awgde_routing}
\centering \Lambda_A=\{\lambda_{An},\cdots,\lambda_{(A+1)n-1}\}
\end{equation}
for $A=0,1,\cdots,n^{d-2}-1$. In the second stage of the network,
the $\gamma$th output of the AWG $A$ and the $A$th input of the Mux
$\gamma$ are connected by an internal link, which carries a stream
of signals with wavelength set $\Lambda_A$. The following path,
denoted as $L(A,a,\gamma)$, from input $(A,a)$ to output $\gamma$:
\setlength{\arraycolsep}{0.1em}
\begin{eqnarray}
\centering (A,a) & \rightarrow & \textrm{the } \gamma\textrm{th
output port
of AWG } A\nonumber\\
 & \rightarrow & \textrm{the } A\textrm{th input port of Mux
}\gamma\nonumber
\end{eqnarray}
is connected through the wavelength
$\lambda_x\in\Lambda_A$, where \setlength{\arraycolsep}{0.1em}
\begin{eqnarray}\label{equ_awgde_wa}
\centering
x &=& An+[a+\gamma]_n\nonumber\\
  &=& \lfloor\alpha/n\rfloor n+\big[[\alpha]_n+\gamma\big]_n.
\end{eqnarray}
Similar to the three-stage decomposition of AWGs described in
\cite{Ye:JLT2012}, the path $L(A,a,\gamma)$ in the two-stage
$\mathcal{N}_I (n^{d-2},n)$ determined by the wavelength routing
(\ref{equ_awgde_wa}) are contention-free at both inputs and outputs,
as illustrated in the proof of the following lemma.

\begin{figure}[t]
\centering
\includegraphics[width=0.49\textwidth]{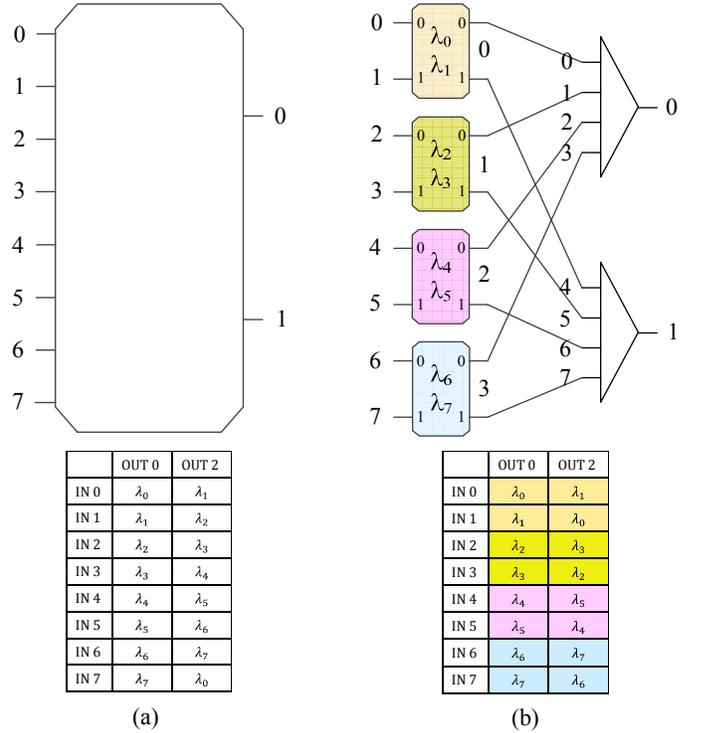}
\caption{Illustration of AWG decomposition: (a) $8\times2$ AWG and
(b) $8\times2$ two-stage AWG network.}\label{f_awg-net-example}
\end{figure}

\begin{lemma}\label{lemma_awgde}
The network $\mathcal{N}_I (n^{d-2},n)$ is functionally equivalent
to an $n^{d-1}\times n$ AWG.
\end{lemma}

\begin{figure*}[t]
\centering
\includegraphics[width=0.6\textwidth]{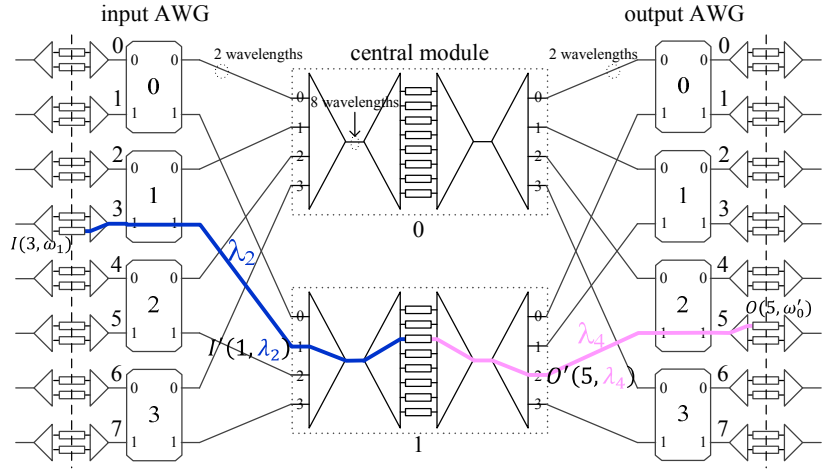}
\caption{Three-stage network
$\mathcal{S}_B(2,2^{4-1},2)$.}\label{f_Sb-example}
\end{figure*}

\begin{figure*}[t]
\centering
\includegraphics[width=0.9\textwidth]{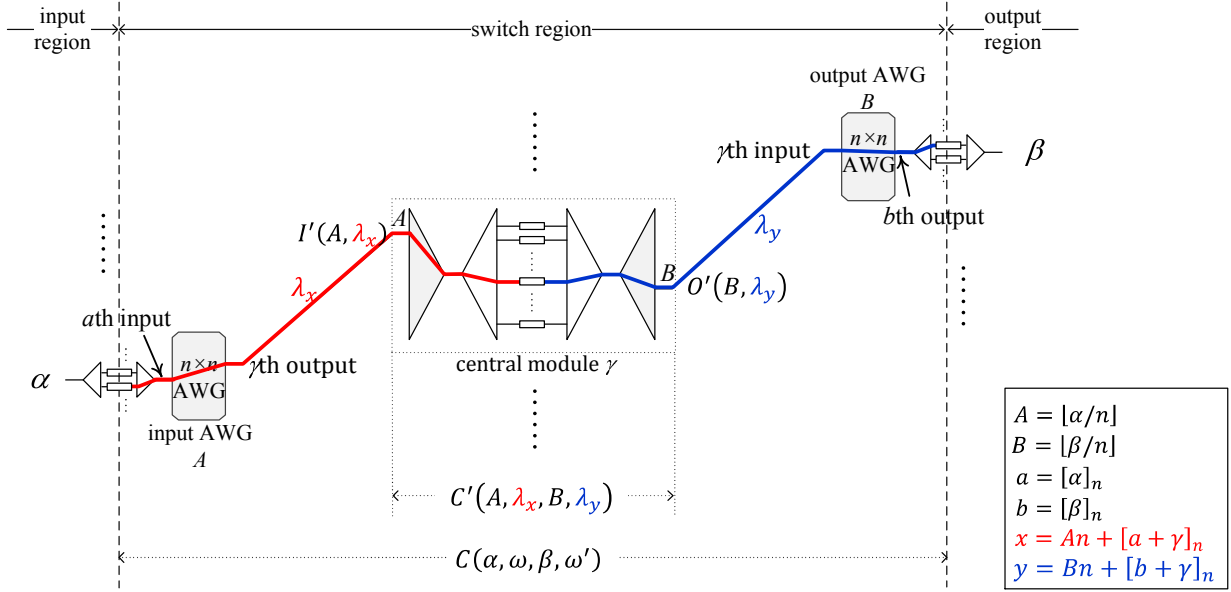}
\caption{Connection of a call $C(\alpha,\omega_i,\beta,\omega_j')$
in the network $\mathcal{S}_B(n,n^{d-1},n)$.}\label{f_Sb}
\end{figure*}

\begin{IEEEproof}
Each input $(A,a)$ and each output $\gamma$ of the network
$\mathcal{N}_I (n^{d-2},n)$ are connected by the path
$L(A,a,\gamma)$. We want to show that these paths are
contention-free at each input and each output if wavelength routing
follows the rule specified by (\ref{equ_awgde_wa}).

Suppose that the two connections $L_1(A_1,a_1,\gamma_1)$ and
$L_2(A_2,a_2,\gamma_2)$ are originated at the same input
$\alpha=(A,a)$, then we have
\[\begin{cases}{}
A_1=A_2=A\\
a_1=a_2=a
\end{cases},\]
If $L_1$ and $L_2$ are routed by the same wavelength
$\lambda_x\in\Lambda_A$, then, from (\ref{equ_awgde_wa}), we have
\[x=An+[a+\gamma_1 ]_n=An+[a+\gamma_2 ]_n,\]
which implies $\gamma_1=\gamma_2$, because
$\gamma_1,\gamma_2=0,1,\cdots,n-1$. It follows that $L_1$ and $L_2$
are the same connection. Thus, all inputs are contention-free.

Similarly, suppose that the two connections $L_1(A_1,a_1,\gamma_1)$
and $L_2(A_2,a_2,\gamma_2)$ are destined for the same output, i.e.,
$\gamma_1=\gamma_2=\gamma$ and use the same wavelength
$\lambda_x\in\Lambda_A$. Then, according to (\ref{equ_awgde_wa}), we
have \[A_1 n+[a_1+\gamma]_n=A_2 n+[a_2+\gamma]_n,\] which implies
$A_1=A_2$, because $[a_1+\gamma]_n,[a_2+\gamma]_n=0,1,\cdots,n-1$.
We therefore have \[[a_1+\gamma]_n=[a_2+\gamma]_n,\] which implies
$a_1=a_2$, because $a_1,a_2=0,1,\cdots,n-1$. Again, the two
connections $L_1$ and $L_2$ must be the same, and all outputs are
also contention-free.
\end{IEEEproof}

The decomposition of an $8\times2$ AWG into the two-stage network
$\mathcal{N}_I(4,2)$ is illustrated in Fig. \ref{f_awg-net-example}.
The network $\mathcal{N}_I(4,2)$ consists of four $2\times2$ AWGs in
the first stage and two $4\times1$ Muxes in the second stage. The
original wavelength set $\Lambda=\{\lambda_0,\cdots,\lambda_7\}$ is
partitioned into four subsets $\Lambda_0=\{\lambda_0,\lambda_1\}$,
$\Lambda_1=\{\lambda_2,\lambda_3\}$,
$\Lambda_2=\{\lambda_4,\lambda_5\}$, and
$\Lambda_3=\{\lambda_6,\lambda_7\}$, associated with the four
corresponding $2\times2$ AWGs.

\begin{figure*}[t]
\centering
\includegraphics[width=0.8\textwidth]{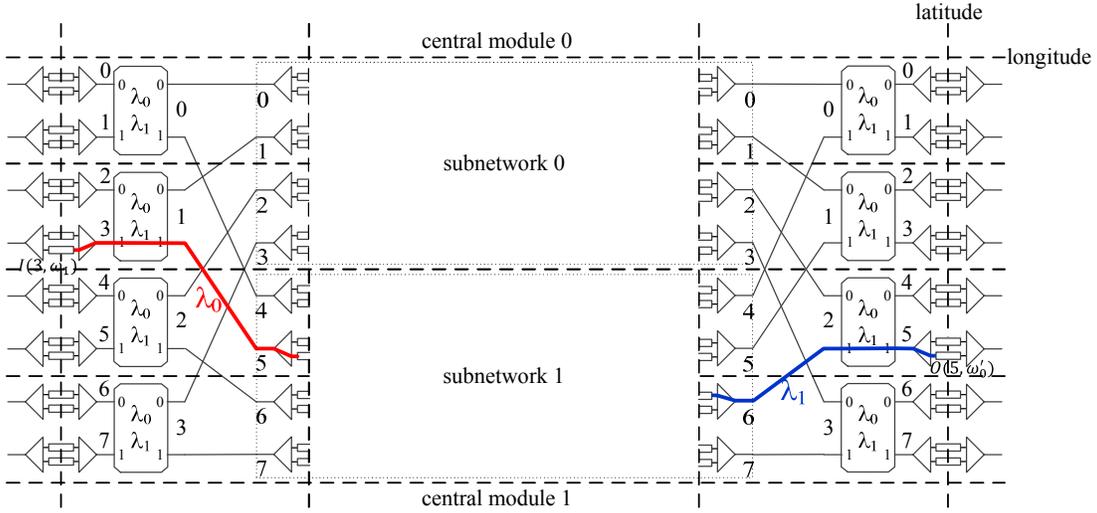}
\caption{Recursive three-stage network
$\mathcal{S}_C(2,2^{4-1},2)$.}\label{f_Sc-example}
\end{figure*}

\begin{figure*}[t]
\centering
\includegraphics[width=0.9\textwidth]{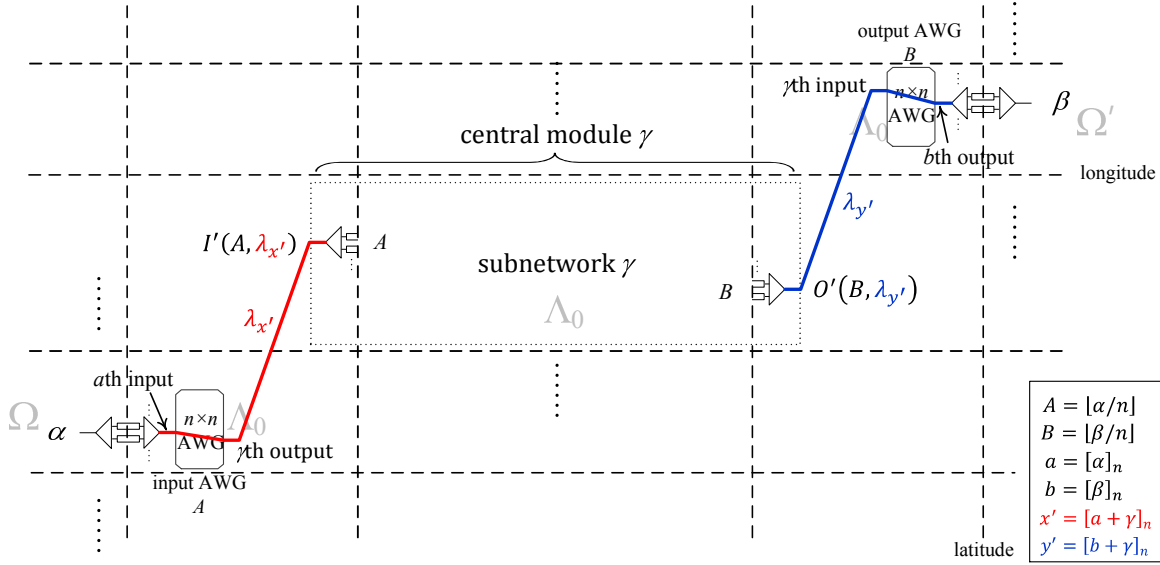}
\caption{Connection of a call $C(\alpha,\omega_i,\beta,\omega_j')$
in the network $\mathcal{S}_C(n,n^{d-1},n)$.}\label{f_Sc}
\end{figure*}

Similarly, it is easy to show that an $n\times n^{d-1}$ AWG can be
decomposed into a functionally equivalent two-stage network
$\mathcal{N}_O(n,n^{d-2})$, which is a mirror image of the network
$\mathcal{N}_I(n^{d-2},n)$. According to Lemma \ref{lemma_awgde}, if
the two AWGs in $\mathcal{S}_A(n,n^{d-1},n)$ are replaced by the
corresponding networks $\mathcal{N}_I(n^{d-2},n)$ and
$\mathcal{N}_O(n^{d-2},n)$, then we obtain an equivalent three-stage
network, denoted as $\mathcal{S}_B(n,n^{d-1},n)$. For example, the
network $\mathcal{S}_B(2,8,2)$ shown in Fig. \ref{f_Sb-example} is
equivalent to the network $\mathcal{S}_A(2,8,2)$ displayed in Fig.
\ref{f_sa-example}. The $n\times n$ AWGs in
$\mathcal{N}_I(n^{d-2},n)$ are referred to as \emph{input AWGs}, and
those in $\mathcal{N}_O(n^{d-2},n)$ are \emph{output AWG}s. In the
network $\mathcal{S}_B(n,n^{d-1},n)$, there are $n$ central modules
and they possess the following properties:
\begin{itemize}
[\IEEEsetlabelwidth{P}]
\item[P1] \emph{There is a unique link between each $n\times n$ AWG and each central
module},
\item[P2] \emph{Each central module is an $n^{d-1}\times n^{d-1}$ WDM switch, which consists of an $n^{d-2}\times1$ Mux, a central TWC-module, and a $1\times n^{d-2}$ DeMux},
\item[P3] \emph{Each central module is functionally equivalent to an $n^{d-1}\times n^{d-1}$ crossbar, where each input or output of the central module carries $n$ wavelengths, and there are $n^{d-1}$ TWC converters with the conversion range of $n^{d-1}$},
\item[P4] \emph{Each central module employs the entire wavelength set $\Lambda=\{\lambda_0,\lambda_1,\cdots,\lambda_{n^{d-1}-1}\}$. According to the non-blocking and contention-free principle, this wavelength set $\Lambda$ is not scalable because the single link in each Mux/DeMux pair carries all wavelengths in $\Lambda$}.
\end{itemize}
Note that the property P4 implies that the wavelength granularity of
the network $\mathcal{S}_B(n,n^{d-1},n)$ is $n^{d-1}$, the same as
that of the network $\mathcal{S}_A(n,n^{d-1},n)$.

\begin{figure*}[t]
\centering
\includegraphics[width=0.99\textwidth]{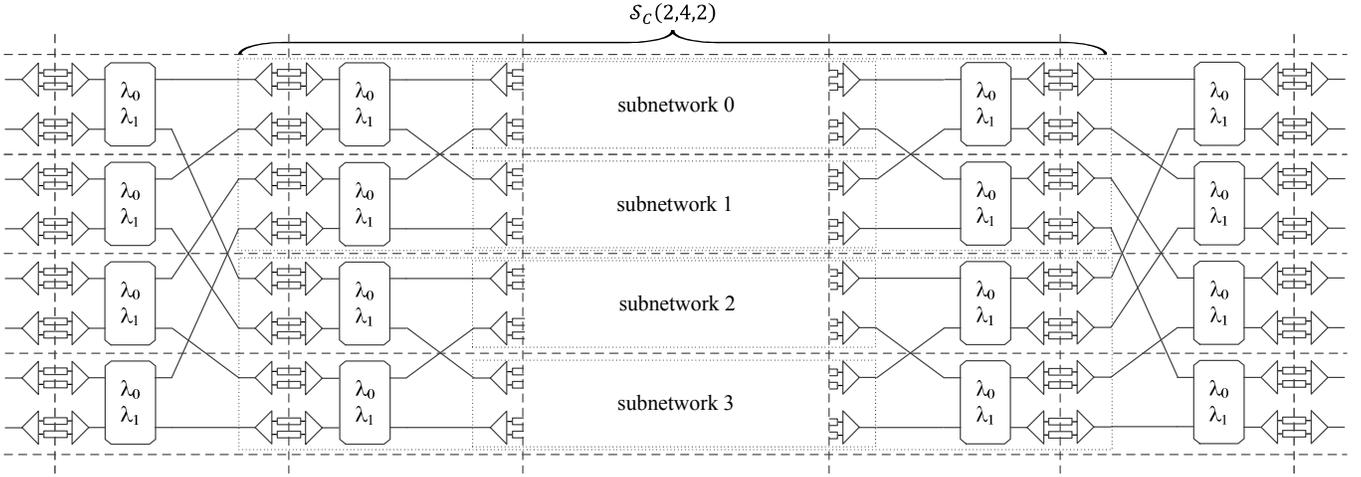}
\caption{Recursive decomposition of the network
$\mathcal{S}_C(2,2^{4-1},2)$.}\label{f_Sc-Sc-example}
\end{figure*}

In the network $\mathcal{S}_B(2,8,2)$ shown in Fig.
\ref{f_Sb-example}, the connection of the call
$C(3,\omega_1,5,\omega_0')$ yields a sub-call from $I'(1,\lambda_2)$
to $O'(2,\lambda_4)$ in the central module 1. Generally, in the
network $\mathcal{S}_B(n,n^{d-1},n)$, a stereotyped connection of
the call $C(\alpha,\omega,\beta,\omega')$ via central module
$\gamma$ will incur a sub-call $C'(A,\lambda_x,B,\lambda_y)$, as
shown in Fig. \ref{f_Sb}. In the central module $\gamma$, the
parameters $A$, $B$, $x$, and $y$ of this sub-call are given by
\begin{equation}\label{equ_Sb_lroute}
\centering A=\lfloor\alpha/n\rfloor,
\end{equation}
\begin{equation}\label{equ_Sb_rroute}
\centering B=\lfloor\beta/n\rfloor,
\end{equation}
\begin{equation}\label{equ_Sb_lwavelength}
\centering x=An+[a+\gamma]_n=An+\big[[\alpha]_n+\gamma\big]_n,
\end{equation}
and
\begin{equation}\label{equ_Sb_rwavelength}
\centering y=Bn+[b+\gamma]_n=Bn+\big[[\beta]_n+\gamma\big]_n.
\end{equation}

Despite the fact that the AWGs are modularized in the network
$\mathcal{S}_B(n,n^{d-1},n)$, it is still not a recursive
decomposition scheme because the central modules do not repeat the
same three-stage structure of the entire network. In addition, the
wavelength set and the conversion range of TWCs are not scalable due
to the properties P3 and P4 above. According to the non-blocking and
contention-free principle, different input (output) AWGs can reuse
the same wavelength set only if they are either topologically
independent or separated by fictitious boundaries of TWCs.

To achieve the wavelength reuse and retain the properties P1 and P2
above, we can logically divide the network into disjoint cells by
the vertical fictitious boundaries in the middle of TWCs, and the
horizontal dashed lines between AWG modules. As Fig.
\ref{f_Sc-example} illustrates, the wavelengths used by the part of
each central module between the two boundaries, referred to as a
\emph{sub-network}, can be independent of those wavelength sets
associated with the AWGs and the TWCs outside these boundaries. If
each sub-network repeats the same three-stage structure of the
entire network, we then have a recursively constructed three-stage
network, denoted as $\mathcal{S}_C(n,n^{d-1},n)$. Moreover, in this
three-stage network, all input and output TWC-modules and all input
and output AWGs can use the same wavelength set, e.g.,
$\Lambda_0=\{\lambda_0,\cdots,\lambda_{n-1}\}$, because they are
either topologically independent or separated by fictitious
boundaries. It follows that the wavelength granularity of the
$\mathcal{S}_C(n,n^{d-1},n)$ is $n-1$, much smaller than that of the
network $\mathcal{S}_A(n,n^{d-1},n)$ and
$\mathcal{S}_B(n,n^{d-1},n)$. For example, every cell in the
$\mathcal{S}_C(2,8,2)$ shown in Fig. \ref{f_Sc-example} uses the
same wavelength set $\Lambda_0=\{\lambda_0,\lambda_1\}$.

The connection of a call in the network $\mathcal{S}_C(n,n^{d-1},n)$
is similar to that in the $\mathcal{S}_B(n,n^{d-1},n)$ described
before. As an example, Fig. \ref{f_Sc-example} illustrates the
connection of the call $C(3,\omega_1,5,\omega_0')$ in the network
$\mathcal{S}_C(2,8,2)$. In general, the connection of a call
$C(\alpha,\omega_i,\beta,\omega_j')$ via central module $\gamma$ in
the network $\mathcal{S}_C(n,n^{d-1},n)$, as Fig. \ref{f_Sc} shows,
can be expressed by
\[I(\alpha,\omega_i)\rightarrow \underset{\textrm{stage 1}}{\underbrace{\alpha}}
\xrightarrow{\lambda_{x'}}\underset{\textrm{stage
2}}{\underbrace{\gamma}}\xrightarrow{\lambda_{y'}}\underset{\textrm{stage
3}}{\underbrace{\beta}}\rightarrow O(\beta,\omega_j'),\] where $x'$
and $y'$ are given as follows:
\begin{equation}\label{equ_Sc_lwavelength}
\centering x'=[a+\gamma]_n=\big[[\alpha]_n+\gamma\big]_n,
\end{equation}
and
\begin{equation}\label{equ_Sc_rwavelength}
\centering y'=[b+\gamma]_n=\big[[\beta]_n+\gamma\big]_n.
\end{equation}
Similarly, in the network $\mathcal{S}_C(n,n^{d-1},n)$, a sub-call
$C'(A,\lambda_{x'},B,\lambda_{y'})$ will be incurred in the central
module $\gamma$ by this connection of the call
$C(\alpha,\omega,\beta,\omega')$.

Collecting the above discussions, we conclude this section in the
following theorem.
\begin{theorem}\label{theorem_Sc}
The network $\mathcal{S}_C(n,n^{d-1},n)$ is an RNB network that
achieves the objective NB1$\sim$NB4.
\end{theorem}
\begin{IEEEproof}
Since the number of central modules in the three-stage network
$\mathcal{S}_C(n,n^{d-1},n)$ is $n$, according to Theorem
\ref{theorem_SA}, it is RNB if every central module is RNB. In
addition, according to Lemma \ref{lemma_awgde}, the wavelength
assignments based on (\ref{equ_Sc_lwavelength}) and
(\ref{equ_Sc_rwavelength}) are contention-free. Next, we want to
show that the objective NB1$\sim$NB4 can be achieved by this
recursive network.

In the network $\mathcal{S}_C(n,n^{d-1},n)$, there are $n^{d-1}$
$n\times n$ input (output) TWC-modules that totally contain $N=n^d$
TWCs, and $n^{d-2}$ $n\times n$ input (output) AWGs. When all
$N=n^d$ wavelength channels are active, all input (output) TWCs and
all input (output) AWGs of the network $\mathcal{S}_C(n,n^{d-1},n)$
are fully occupied. Also, as we discussed above, all TWCs and AWGs
are associated with the same wavelength set $\Lambda_0$, as Fig.
\ref{f_Sc} illustrates. Thus, the first and third stages of the
network $\mathcal{S}_C(n,n^{d-1},n)$ fulfill the objective
NB1$\sim$NB4. Moreover, each sub-network in the central module
repeats the same three-stage structure of the entire network, and
the wavelengths used by the sub-networks are independent of those
wavelength sets associated with the TWCs and AWGs outside these
boundaries. It follows that all $n$ central modules can also fulfill
these objectives if we repeatedly apply the recursive decomposition
scheme $\mathcal{S}_C(n,n^{d-1},n)$ to each central module.
\end{IEEEproof}

\begin{figure*}[t]
\centering \subfigure[]{
\includegraphics[width=0.99\textwidth]{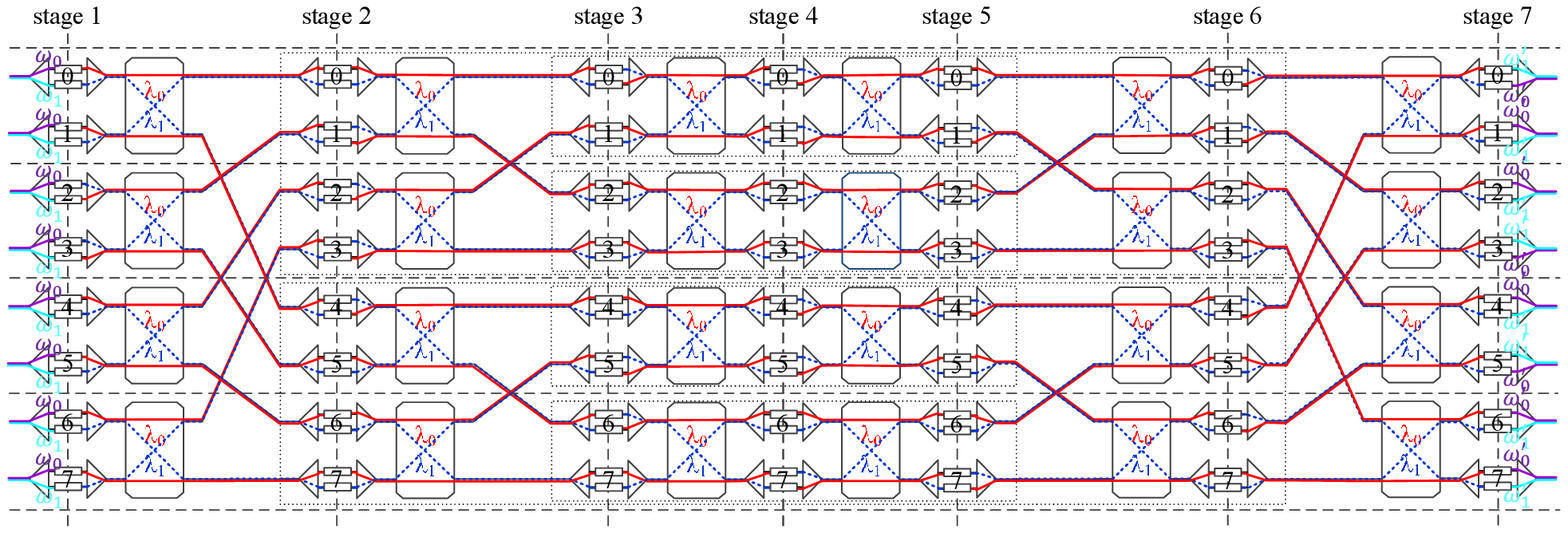}}
\subfigure[]{
\includegraphics[width=0.9\textwidth]{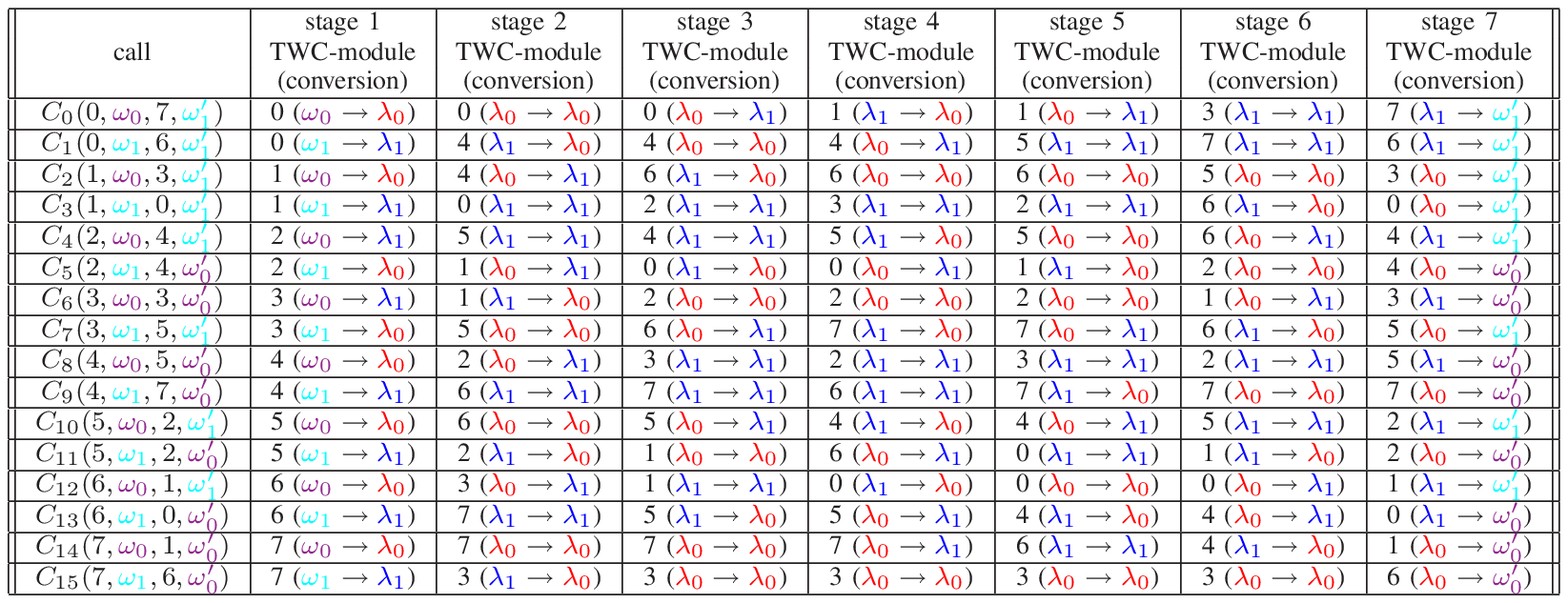}}
\caption{A fully loaded AWG-based three-stage network: (a)
$\mathcal{B}(2,4)$ and (b) the route and wavelength
assignment.
}\label{f_b-example}
\end{figure*}

An example is illustrated in Fig. \ref{f_Sc-Sc-example}, which is
obtained by the decomposition of the central modules of
$\mathcal{S}_C(2,8,2)$ network shown in Fig. \ref{f_Sc-example}. It
is clear that all TWC-modules and AWG modules can be associated with
the same wavelength set $\Lambda_0$, and fulfill the objective
NB1$\sim$NB4.

Repeatedly applying the recursive scheme, a WDM network with $N=n^d$
input and output channels can be constructed by $n\times n$
TWC-modules and $n\times n$ AWGs associated with a set of $n$
wavelengths $\Lambda_0=\{\lambda_0,\cdots,\lambda_{n-1}\}$. In
particular, if we apply the decomposition scheme $d-2$ times to the
network $\mathcal{S}_A(n,n^{d-1},n)$, then we obtain an AWG-based
three-stage network denoted as $\mathcal{B}(n,d)$, which consists of
$2d-1$ columns of $n\times n$ TWC-modules and $2d-2$ columns of
$n\times n$ AWGs. Clearly, the wavelength granularity of the network
$\mathcal{B}(n,d)$ is $n$, which is only $1/ n^{d-2}$ of that of the
network $\mathcal{S}_A(n,n^{d-1},n)$. For example, the network shown
in Fig. \ref{f_b-example}(a) is the AWG-based three-stage network
$\mathcal{B}(2,4)$ with wavelength granularity 2. The route and
wavelength assignments tabulated in Fig. \ref{f_b-example}(b)
illustrate that the AWG-based three-stage network can achieve 100\%
utilization when all input wavelength channels are busy.

\begin{figure}[t]
\centering
\includegraphics[width=0.49\textwidth]{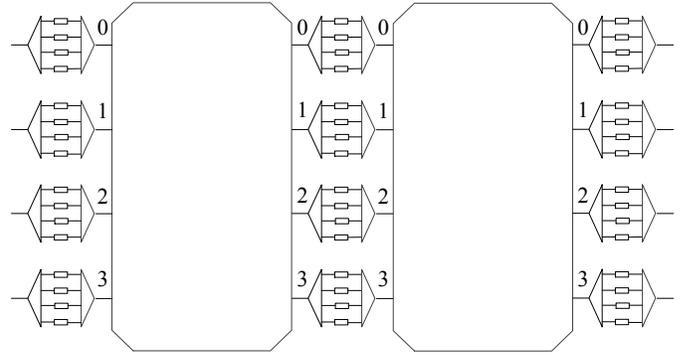}
\caption{An AWG-based three-stage network $\mathcal{B}(4,2)$.
}\label{f_b4}
\end{figure}

Therefore, an $N\times N$ AWG-based three-stage network $\mathcal{B}(n,d)$
can be recursively constructed from a set of $n\times n$ AWGs
together with a collection of TWCs with conversion range $n$, where
$N=n^d$. There is an inherent tradeoff between the physical
interconnection complexity and the wavelength granularity in the
AWG-based WDM switches, and this point can be demonstrated by a
comparison between the network $\mathcal{B}(2,4)$ shown in Fig.
\ref{f_b-example}(a) and the network $\mathcal{B}(4,2)$ displayed in
Fig. \ref{f_b4}. The number of interconnection links between two
adjacent stages in an $N\times N$ electronic three-stage network is $N$ or
$n^d$, while the counterpart in an AWG-based three-stage network
$\mathcal{B}(n,d)$ is $n^{d-1}$, because each link in
$\mathcal{B}(n,d)$ can simultaneously carry $n$ wavelength channels.
This reduction factor $n$ of interconnection complexity can be
significant if $n$ is large, which is achieved at the expense of
increasing the size of AWG modules and the conversion range of TWCs.

\begin{table*}[t]
\centering \caption{Comparison of different AWG-based switching
networks}
\begin{tabular}{||c|c|c|c|c|c||}
\hline  & NB1 & NB2 & NB3 & NB4 & physical\\
\rb{Schemes} & (100\% utilization) & (modular AWG) & (scalable TWC) & (wavelength reuse) & interconnection complexity\\
\hline AWG-based Crossbar &  &  &  &  & \\
\cite{Blumenthal:JSTQE,Bianco:TON,Cheyns:PNC2003,Pan:JLT2005,Lucerna:HPSR2011,Maier:SPIE2000} & \rb{no} & \rb{no} & \rb{no} & \rb{no} & \rb{$O(1)$} \\
\hline Single-stage network &  &  &  &  & \\
\cite{Pattavina:INFOCOM2006} & \rb{no} & \rb{no} & \rb{no} & \rb{no} & \rb{$O(N)$} \\
\hline Multi-stage network &  &  &  &  & \\
\cite{Cheyns:CM2004,Ngo:TON2006,Zanzottera:HPSR2006} & \rb{no} & \rb{yes} & \rb{yes} & \rb{yes} & \rb{$O(N)$} \\
\hline $\mathcal{S}_A(n,r,m)$ &  &  &  &  & \\
\cite{Zhong:JLT1996,Leonardus:EU2007}, Theorem \ref{theorem_SA} & \rb{yes} & \rb{no} & \rb{no} & \rb{no} & \rb{$O(1)$} \\
\hline $\mathcal{S}_B(n,r,m)$ & yes & yes & no & no & $O(N/n)$ \\
\hline $\mathcal{B}(n,d)$, Theorem \ref{theorem_Sc} & yes & yes & yes & yes & $O(N/n)$ \\
\hline
\end{tabular}\label{t_comparison}
\end{table*}

In the practical implementation of the AWG-based three-stage network
$\mathcal{B}(n,d)$, due to current state of art, we should take the following limitations into consideration:
\begin{itemize}
\item[1)] Coherent crosstalk: Let $p$ (in dB) be the power penalty induced
by the coherent crosstalk of an $n\times n$ AWG when all input
wavelengths are the same. Since there are $2\log_n N$ stages of
$n\times n$ AWGs in an $N\times N$ $\mathcal{B}(n,d)$ network, the
overall power penalty of each connection is $2p\log_n N$ if there is
no compensation mechanism within the network;
\item[2)] Amplified spontaneous emission (ASE) noise:
Semiconductor optical amplifiers (SOAs) are the key components of a
compact all-optical TWC. The ASE noise generated by SOAs will
decrease the optical signal-to-noise ratio (OSNR). Therefore, if
SOA-based all-optical TWCs without regenerative function are used,
then the OSNR will be faded with the increasing number of stages;
\item[3)] Insertion loss: A typical AWG has a $\sim$6-dB insertion loss.
\end{itemize}
These limitations could be relaxed as follows. First, keep $n$
smaller than 15 \cite{Gaudino:ICC2008} such that the coherent
crosstalk can be compensated at the outputs of each $n\times n$ AWG.
Second, use all-optical regenerative TWCs
\cite{Chayet:OFC2004,Porzi:PTL2013} or optic-electron-optic (O-E-O)
TWCs \cite{Blumenthal:JSTQE} in the network such that 3R signal
regenerations can be performed stage by stage. In practice, it is
feasible to use O-E-O TWCs now because high-speed ($\sim$100 GHz)
transceivers are already commercially available.

With the maturity of photonics integrated circuits, the modular
construction of an optical switch by using optical integrated
modules is gradually workable. For example, the $8\times 8$
AWG-based crossbar on a III/V photonics chip reported in
\cite{Blumenthal:JSTQE} demonstrates the feasibility of the
integration of each stage of the $\mathcal{B}(n,d)$ network on a
single photonics chip. However, it is still difficult to integrate
the entire optical switch network, including the optical switching
fabric and the electronic controller, on a single chip based on
silicon-on-insulate (SOI) technology, despite the fact that the silicon-based
AWGs \cite{Cheung:CLEO2012}, and the silicon-based modulator and
detector \cite{Dhiman:JAP2013}, the key component of the O-E-O TWC,
are both commercially available. The bottleneck is the unexplored
technique of a silicon-based optical laser, an essential component of
the TWC. Another challenge of the integration on a single chip is that of
the crosspoints of the interconnection links between two adjacent
stages, since additional crosstalk and loss could be incurred if the
number of crosspoints is too large.

\section{Conclusion}\label{conclusion}
This paper proposes a recursive scheme for the construction of
AWG-based non-blocking Clos networks. In respect to the network
scalability and optimal utilization criteria, this recursive
approach outperforms the previous schemes in the following aspects.
First, it can construct a large-scale network by a collection of
small-sized AWGs while preserving the complete wavelength routing
property. Second, it logically divides the WDM switch network into
wavelength independent cells, such that a similar, but smaller, set
of wavelengths can be reused by these cells, which substantially
scales down the wavelength granularity and the conversion range of
TWCs. Third, the route assignments in these recursive networks are
consistent with those in the classical Clos networks, and thus the
optimal AWG-based RNB Clos network can achieve 100\% utilization
when all input and output wavelength channels are busy. For
comparison purposes, Table \ref{t_comparison} lists all known
results on the AWG-based multistage networks.

\bibliography{IEEEabrv,Bib_YT}

\end{document}